\documentstyle[pra,aps,epsfig,amsmath,amssymb]{revtex}
\twocolumn
\begin{document}
\draft
\title{The role of the initial state in the collapse of a Bose condensed gas}
\author{Thorsten K\"ohler}
\address{Clarendon Laboratory, Department of Physics, 
University of Oxford, Oxford, OX1 3PU, United Kingdom}
\date{\today}
\maketitle
\begin{abstract}
The stability of a Bose condensed gas with a negative scattering length 
is studied with regard to the role played by the initial state.  
A trapped ideal gas ground state is shown to be unstable when the 
Gross Pitaevskii equation still predicts the 
existence of a stable state. A possible relation to a recent 
experimental study of the critical conditions for stability or collapse 
of a Bose Einstein condensate 
[J.~L.~Roberts et al., Phys.~Rev.~Lett.~{\bf 86}, 4211 (2001)]
is discussed.
\end{abstract}

\pacs{PACS numbers: 03.75.Fi, 34.50.-s}

The possibility of manipulating interatomic forces using magnetic field
tunable Feshbach resonances 
\cite{Inouye},
has provided new opportunities for the study
of collision phenomena in ultra cold gases. Recent 
experiments with $^{85}$Rb 
\cite{Cornish}
have shown how one can tune the two body 
interaction, in a condensate from strongly repulsive to  
attractive over a wide range of different $s$ wave scattering 
lengths. This makes possible the study of non equilibrium phenomena
under a variety of controlled initial conditions.
One of 
the most striking experiments of this type is that in which a condensate,
first formed with repulsive forces, is switched to the attractive case,
and collapses.
In an idealized sense collapse means an unrestricted contraction of a gas 
within a finite amount of time. As shown by Pitaevskii 
\cite{Pitaevskii},
the time dependent Gross Pitaevskii 
equation (GPE) predicts this remarkable phenomenon for a trapped
Bose condensed gas with a negative scattering length in a harmonic 
trap as soon as certain 
conditions, on the interaction strength, and the initial condensate 
wave function, are fulfilled. 

Very recently, the control of the interatomic
force in $^{85}$Rb \cite{Roberts}
has allowed an experimental determination of the 
critical conditions for the stability, with respect to collapse, 
of a condensate. In the course of these experiments, the 
interatomic force of the gas was tuned towards
specific negative $s$ wave scattering lengths by means of magnetic 
fields, using a Feshbach resonance. Since the initial scattering
length was chosen to be negligibly small, from the point of view of the 
GPE the prepared initial state can be interpreted as an ideal gas 
condensate wave function.

Since their first observation in experiments with $^7$Li 
\cite{Bradley},
Bose condensed gases with a negative scattering length have been studied
by numerous authors. Most of this work has used the GPE and focused on the 
stability conditions for a ground state solution
(see, e.g., \cite{Dalfovo} and references therein)
or condensate formation
\cite{Sackett}.
The experimental results in Ref.~\cite{Roberts} indicate, however, that
as soon as the condensate becomes unstable the GPE alone is not suitable 
for the description of the dynamics of the collapse. In particular
the observed substantial loss of condensate atoms contradicts the number 
conservation inherent in the GPE.
This failure can be expected because the collision term of the GPE
does not account for collisional losses of condensate atoms due to 
non forward scattering 
\cite{Band},
which should become relevant for the rapid dynamics occurring
during the collapse.

More interestingly the gas was seen to be unstable
even under conditions, where the GPE predicts the existence of a 
stable ground state. To clarify this aspect of the observations 
we have studied the 
stability of a Bose condensed gas with a negative scattering length
using the GPE, taking into account the role of the  
initial state, prepared in the experiment in Ref.~\cite{Roberts}.
We show that      
this initial ideal gas state leads to collapse condition that is different
from that for the existence of a ground state of the condensate.

Throughout this article the $s$ wave scattering length of the
interatomic force, $a_0$, is assumed to be negative. 
For a Bose condensed gas in a spherical trap, with the harmonic
potential $V_{\rm trap}=m\omega_{\rm ho}^2 r^2/2$, the natural units
of time and length are the inverse trap frequency 
$\omega_{\rm ho}^{-1}$ and the harmonic oscillator length 
$a_{\rm ho}=\sqrt{\hbar/m\omega_{\rm ho}}$, respectively. 
Expressed through these units the GPE 
for an interacting Bose gas, consisting of $N$ atoms, 
assumes the well known form
\cite{Dalfovo}
\begin{equation}
i\frac{\partial}{\partial t}\psi=
-\frac{1}{2}\Delta\psi+\frac{1}{2}r^2\psi-
4\pi k|\psi|^2\psi.
\label{GPE}
\end{equation}
Here the condensate wave function $\psi$ is normalized to unity,
$k$ is given by $N|a_0|/a_{\rm ho}$ and 
the corresponding unit of the energy is then
$N\hbar^2/ma_{\rm ho}^2$. Equation (\ref{GPE})
shows that $k$ is the only free adjustable parameter of the GPE 
for a spherical trap. 

On first sight, the form of Eq.~(\ref{GPE}) might suggest 
that the critical conditions for collapse are governed 
solely by the parameter $k$. With respect to the existence of a stable 
ground state of a condensate this is true. In fact, the  
theoretical studies in Ref.~\cite{Ruprecht,Eleftheriou}
confirm the condition for stability
\begin{equation} 
k<k_{\rm cr}=0.575.
\label{standardcriticalcond}
\end{equation} 
When the initial condensate wave function deviates from the ground state,
however, a collapse can occur even if the actual parameter $k$
is well below the above standard value. 

As proposed by Pitaevskii 
\cite{Pitaevskii}
the conditions for stability or collapse can be visualized by means
of two fundamental estimates provided by the GPE. The first one is
the uncertainty relation between the mean square of the radius of a 
condensate, $\langle r^2\rangle=\int d{\bf r} r^2 |\psi|^2$, and its kinetic 
energy, $T=\int d{\bf r}|\bigtriangledown\psi|^2/2$, which reads
\begin{equation}
\langle r^2\rangle T\geq 9/8,
\label{est1}
\end{equation}
where the choice of units corresponds to Eq.~(\ref{GPE}). 
It is worth noting that once $\langle r^2\rangle$ vanishes the condensate 
wave function has collapsed. The kinetic energy then becomes infinite.
The second estimate consists in an inequality between the 
mean field energy, $E_{\rm mf}=-4\pi k\int d{\bf r}|\psi|^4/2$, and the 
kinetic energy,
\begin{equation}
\int d{\bf r}|\psi|^4/2\leq \beta T^{3/2},
\label{est2}
\end{equation} 
where $\beta=0.0575$ is a universal constant. Equations (\ref{est1})
and (\ref{est2}) are preserved for all times by the GPE. 

\bigskip
\bigskip

\begin{figure}[htb]
\begin{center}
\epsfig{file=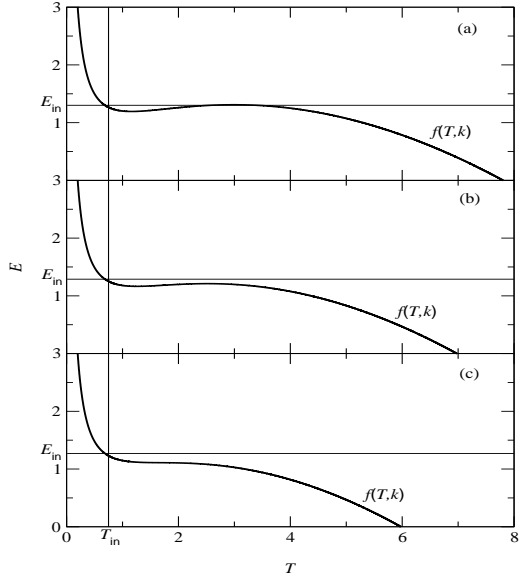,width=8.5cm,height=9cm,angle=0}
\end{center}
\caption{Lower estimate for the total energy $E$ as a function
of the kinetic energy $T$ for three different parameters
of the interatomic force: (a) $k=0.5$, (b) $k=0.53$ and 
(c) $k=0.575$. The horizontal lines indicate the total energy for an
initial ideal gas ground state condensate wave function. The vertical 
line indicates the initial kinetic energy of the Bose condensed gas. 
The energies are given in units of $N\hbar^2/ma_{\rm ho}^2$. The 
accessible points $(T,E)$ are above the curve $f(T,k)$.
\label{Fig:stability}}
\end{figure}

As an additional constant of motion, the total energy, 
\begin{equation}
E=T+V+E_{\rm mf},
\label{energy}
\end{equation}
contains the kinetic and the mean field part as well as the trap energy
$V=\int d{\bf r}V_{\rm trap}|\psi|^2$. In the spherical case the mean 
square of the radius of a condensate and the trap energy are related through
\begin{equation}
V=\langle r^2\rangle/2.
\label{trapenergy}
\end{equation}
The inequality given by Eq.~(\ref{est1}) then provides an estimate for 
the trap energy Eq.~(\ref{trapenergy}) through the kinetic energy.
In addition, according to Eq.~(\ref{est2}), the mean field contribution 
to the total energy Eq.~(\ref{energy}) can be estimated in terms of $T$.
This in turn implies the general relation 
\begin{equation}
E\geq T+\frac{9}{16}T^{-1}-4\pi\beta kT^{3/2}\equiv f(T,k),
\label{est}
\end{equation} 
which is valid for all times.

In Fig.~\ref{Fig:stability} the lower bound Eq.~(\ref{est}) is illustrated
for the parameters $k=0.5$ (Fig.~\ref{Fig:stability}a), 
$k=0.53$ (Fig.~\ref{Fig:stability}b) and 
$k=0.575$ (Fig.~\ref{Fig:stability}c) and an initial
ideal gas ground state. For this Gaussian initial condensate wave function 
the total energy and the initial kinetic energy are given by 
$E=E_{\rm in}=3/2-k/\sqrt{2\pi}$ and $T_{\rm in}=3/4$, respectively.
In Fig.~\ref{Fig:stability}a the function $f(T,k)$ 
exhibits a local minimum which  
encloses the point $(T_{\rm in},E)$. As the total energy is a constant
of motion and $T(t)$ varies continuously in time the kinetic energy 
is caught in the valley for all times. As pointed out by Pitaevskii
\cite{Pitaevskii},
the inequality given by Eq.~(\ref{est1}) then predicts the mean square 
of the radius of the condensate to be positive, 
i.e.~$\langle r^2\rangle \geq 9/8T>0$, and a total collapse, 
i.e.~$\langle r^2\rangle=0$, can never occur. 
As can be seen in Fig.~\ref{Fig:stability}c, very close to the standard
critical interaction strength Eq.~(\ref{standardcriticalcond}) 
the function $f(T,k)$ 
becomes monotonous. Hence, a ground state of the GPE ceases to exist as 
soon as the local minimum of $f(T,k)$ vanishes.   

In Fig.~\ref{Fig:stability}b, however, the total energy is above
the local maximum of $f(T,k)$ and at least the lower bound in 
Eq.~(\ref{est}) allows the kinetic energy to access arbitrarily
high values. In this case a total collapse of the condensate wave
function cannot be excluded by the inequality Eq.~(\ref{est1}).  

\begin{figure}[htb]
\begin{center}
\epsfig{file=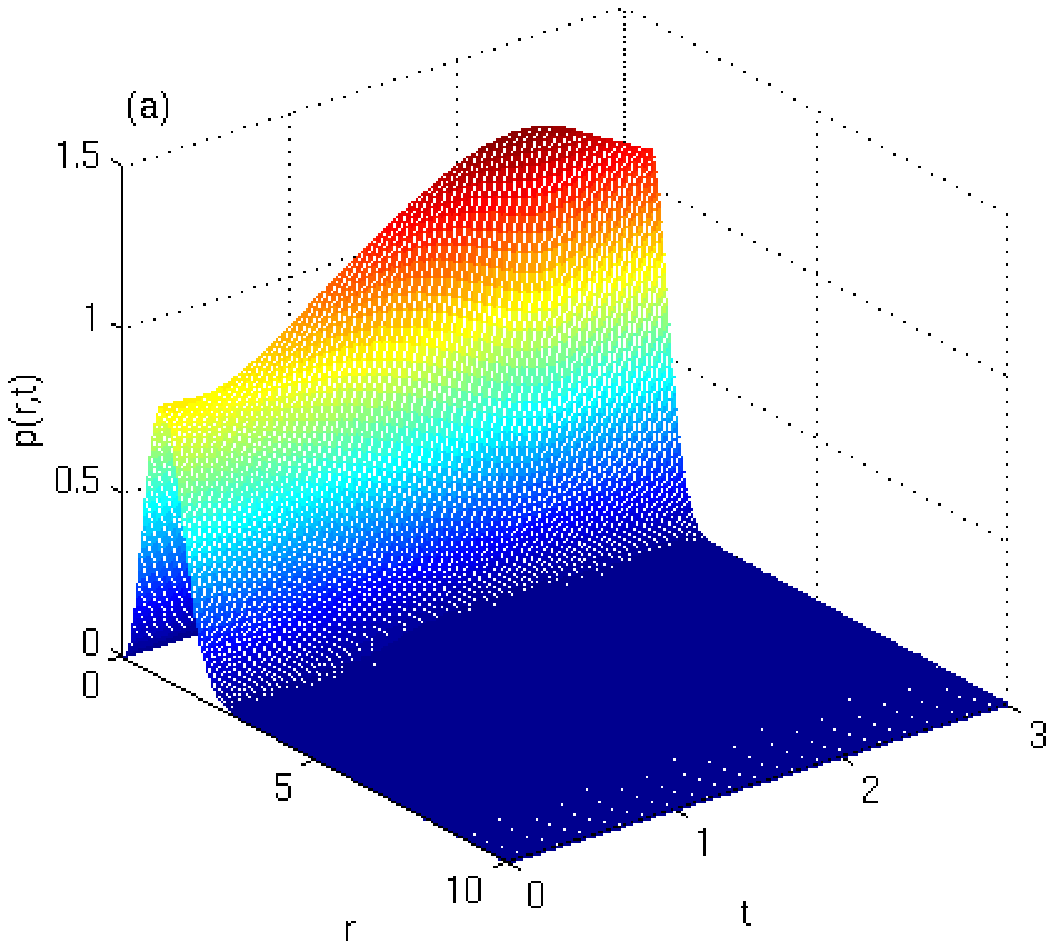,width=8.5cm,height=8.5cm,angle=0}
\epsfig{file=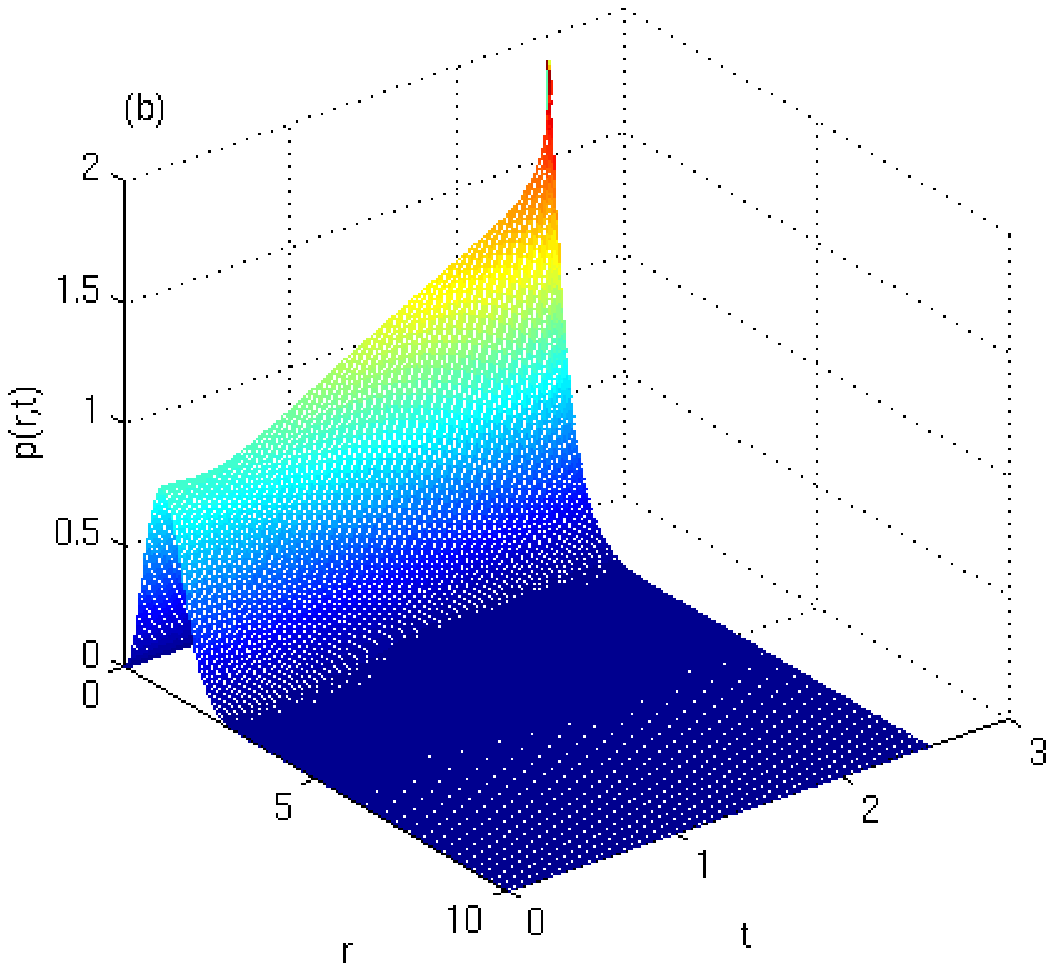,width=8.5cm,height=8.5cm,angle=0}
\end{center}
\caption{Numerical solutions of the time dependent
Gross Pitaevskii equation for a spherical trap with an initial
ideal gas ground state condensate wave function and different 
parameters of the interatomic force:
(a) $k=0.52$ and (b) $k=0.53$. The graphs show the
radial density $p(r,t)=4\pi r^2|\psi(r,t)|^2$ as a function of the radius
$r$ and the time $t$. The time and the radius are given in units of
the inverse trap frequency $\omega_{\rm ho}^{-1}$ and the harmonic
oscillator length $a_{\rm ho}=\sqrt{\hbar/m\omega_{\rm ho}}$, 
respectively. The radial density is normalized to unity.
\label{Fig:simulations}}
\end{figure}

To clarify this situation, Fig.~\ref{Fig:simulations} shows
the results of 
numerical simulations of the GPE in Eq.~(\ref{GPE}) for 
$k=0.52$ (Fig.~\ref{Fig:simulations}a) and 
$k=0.53$ (Fig.~\ref{Fig:simulations}b), with an initial 
ideal gas ground state condensate wave function. 
In Fig.~\ref{Fig:simulations}a, the
simulation predicts the condensate to be stable with respect to collapse
within the considered period of time. 
For the present parameter $k=0.52$ the total energy is already slightly
above the local maximum of the function $f(T,k)$ and a
rigorous assertion concerning stability or collapse
cannot be obtained directly from Pitaevskii's estimate.

\bigskip

\begin{figure}[htb]
\begin{center}
\epsfig{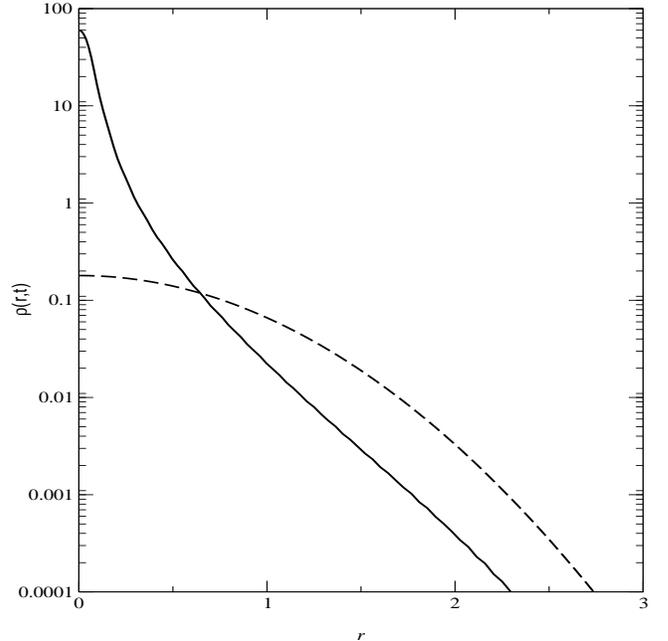}
\end{center}
\caption{Comparison of the spatial condensate density 
$\rho(r,t)=|\psi(r,t)|^2$ at time
$t=2.52\ \omega_{\rm ho}^{-1}$, close to the collapse, (solid line) 
with the initial Gaussian density (dashed line)
for $k=0.53$. The unit of the radius $r$ is chosen as in 
Fig.~\protect\ref{Fig:simulations}. The normalized densities
($\int d{\bf r}\rho(r,t)=1$) are shown on a logarithmic 
scale.\label{Fig:collapse}}
\end{figure}

In Fig.~\ref{Fig:simulations}b, however,
the condensate wave function exhibits a collapse although the actual 
parameter of the interatomic force of $k=0.53$ is still sufficiently 
weak for a stable ground state of the GPE to exist. 
To visualize the contraction of the condensate wave function, in 
Fig.~\ref{Fig:collapse} a typical density $|\psi(r,t)|^2$, close to the 
time of the collapse, is compared with the initial Gaussian density. 
The corresponding wave function is obtained from the simulation in 
Fig.~\ref{Fig:simulations}b. As can be seen on the logarithmic scale,
the predicted condensate density, close to the collapse, deviates from a
Gaussian and exhibits a sharp peak at its center.

The GPE thus predicts the preparation of an initial ideal gas ground state 
condensate wave function to lead to a pronounced reduction 
of the critical parameter of 
the interatomic force to
\begin{equation} 
k_{\rm cr}\lesssim 0.53
\label{criticalupper}
\end{equation} 
in comparison 
with the standard critical value for the existence of a stable ground state. 
This critical value for an initial ideal gas ground state has been 
reported also in Ref.~\cite{Eleftheriou} as an intermediate result in 
the numerical study of stable ground states of the GPE. A numerical
simulation alone, however, cannot predict stability of a condensate on 
arbitrarily long time scales. The required lower estimate of $k_{\rm cr}$ 
is thus provided by the analysis illustrated in Fig.~\ref{Fig:stability}
and yields
\begin{equation} 
k_{\rm cr} > 0.5.
\label{criticallower}
\end{equation}

The sharply defined critical parameter of 
\begin{equation}
k_{\rm cr}=0.461\pm 0.012\pm0.054,
\end{equation} 
reported in Ref.~\cite{Roberts},
is indeed significantly smaller 
than the predicted standard value for the existence of a stable ground 
state in Eq.~(\ref{standardcriticalcond}). 
This corroborates the predictions of the GPE in view of the
destabilizing role of the particular initial state that was prepared 
in the experiment.
Equations (\ref{criticalupper}) and (\ref{criticallower}), however, are not
strictly valid in the experimental case because the magnetic field was  
ramped linearly to its final value rather than abruptly. 
Interestingly, though, our result obtained from the GPE agrees with
the observations  
within the experimental upper limit for $k_{\rm cr}$. 

The analysis of stability or collapse of a dilute Bose condensed 
gas with a negative scattering length, performed in this work, 
can be applied to 
arbitrary initial condensate wave functions and trap geometries. The 
optimal estimate from Eq.~(\ref{est}) is obtained for a spherical trap.
Related experiments with more general initial states should be feasible 
with the method described in Refs.~\cite{Cornish,Roberts} and could provide 
a further understanding of the range of validity of the GPE in non 
equilibrium situations.  

\acknowledgements
The author thanks Thomas Gasenzer,
Samuel Morgan and Keith Burnett for stimulating
discussions. This work was supported by the Alexander von Humboldt 
Foundation.

\end{document}